\documentclass[twocolumn,showpacs,preprintnumbers,amsmath,amssymb]{revtex4}
\usepackage{graphicx}% Include figure files
\usepackage[ansinew]{inputenc}

\begin{document}

\renewcommand{\vec}[1]{\ensuremath{\boldsymbol{#1}}}
\newcommand{\bra}[1]{\ensuremath{\langle #1|}}   %defines a bra
\newcommand{\ket}[1]{\ensuremath{| #1\rangle}}   %defines a ket
\newcommand{\qd}{\ket{\mbox{$\downarrow$}}}
\newcommand{\qu}{\ket{\mbox{$\uparrow$}}}

\title{Optimal control of entangling operations for trapped ion quantum computing}

\author{V. Nebendahl$^{1}$}
\author{H. H{\"a}ffner$^{1,2}$}
\author{C. F. Roos$^{1,2}$}
\email{Christian.Roos@uibk.ac.at}

\affiliation{$^1$Institut f\"ur Quantenoptik und
Quanteninformation, \"Osterreichische Akademie der Wissenschaften,
Otto-Hittmair-Platz 1, A-6020 Innsbruck, Austria}
\affiliation{$^2$Institut f\"ur Experimentalphysik, Universit\"at
Innsbruck, Technikerstr.~25, A-6020 Innsbruck, Austria}

\date{\today}% It is always \today

\begin{abstract}
Optimal control techniques are applied for the decomposition of
unitary quantum operations into a sequence of single-qubit gates
and entangling operations. To this end, we modify a
gradient-ascent algorithm developed for systems of coupled nuclear
spins in molecules to make it suitable for trapped ion quantum
computing. We decompose unitary operations into entangling gates
that are based on a nonlinear collective spin operator and
complemented by global spin flip and local light shift gates.
Among others, we provide explicit decompositions of controlled-NOT
and Toffoli gates, and a simple quantum error correction protocol.
\end{abstract}

% PACS numbers
                 % 32.80.Pj: Opt.cooling of atoms;trapping  (obsolete)
                 % 37.10.Ty: Ion trapping
                 % 32.80.Qk: Coherent control of atomic interactions with photons
                 % Quantum information, 03.67.–a

\maketitle

\section{\label{sec:intro}Introduction}
Choosing the laws of quantum physics as the physical basis for
constructing models of computation \cite{Deutsch:1985} allows for
solving certain computational problems more efficiently as in
models based on classical physics \cite{Nielsen2000}. In the
quantum circuit model, information is encoded in quantum bits
(qubits) and manipulated by applying unitary operations acting on
the joint state space of the qubits. It has been shown that
arbitrary unitary operations can be broken down into sequences of
elementary gate operations, consisting of single-qubit operations
and entangling operations acting on pairs of qubits
\cite{Barenco1995}. It is a non-trivial task to find optimum
decompositions of unitary operations into a minimum number of
elementary gates. Often, the controlled-NOT (CNOT) gate operation
is chosen as the entangling operation. However, it has been shown
that almost any entangling operation can be used for this purpose
as well \cite{Lloyd1995}.

In experiments processing quantum information, the available
physical processes determine the choice of the entangling gate. In
the case of trapped ions manipulated by coherent laser light
\cite{Blatt:2008}, a qubit is realized by encoding quantum
information in a pair of long-lived internal states, consisting of
hyperfine or Zeeman ground states, or in a combination of a ground
state and a metastable state with an energy difference of a few
electron volts. Gates acting on a single qubit are achieved by
lasers coupling the qubit states by either dipole-forbidden
single-photon transitions or Raman transitions. Pairs of ions are
entangled by qubit-qubit interactions mediated by a coupling to a
vibrational mode of the ions' motion in the trap.

To achieve a gate on either a specific ion or a pair of ions
within an ion string, two strategies are pursued:
\begin{enumerate}
\item The laser beam is tightly focussed so that it interacts only
with a single ion at a time \cite{Naegerl:1999}. Then, a CNOT gate
or an equivalent entangling gate between an arbitrary pair of ions
is achieved by a sequence of pulses with the laser addressed to
either one or the other ion of the pair \cite{Schmidt-Kaler2003a}.
\item Alternatively, a laser with wider beam diameter is employed
in combination with ions held in a segmented ion trap. In this
approach, trap potentials are dynamically transformed to enable
the transport of one or two ions into the interaction region with
the laser beam. To entangle a pair of ions $i,j$, a bichromatic
laser field is used to realize either a conditional phase gate
\cite{Leibfried2003a} induced by an effective Hamiltonian
$H_{PG}\propto\sigma_z^{(i)}\sigma_z^{(j)}$ or a
M{\o}lmer-S{\o}rensen gate \cite{Sackett2000} induced by
$H_{MS}\propto\sigma_x^{(i)}\sigma_x^{(j)}$ where $\sigma_n^{(k)}$
denotes a Pauli spin operator $\mathbf{\boldsymbol{\sigma}\cdot
n}$ acting on the $k$'th ion.
\end{enumerate}
Recently, the latter interaction has been used to entangle a pair
of ions with fidelities of up to 99.3(1)\%, with a coupling
mediated by the longitudinal center-of-mass mode of the ion string
\cite{Benhelm:2008b}. Employing the same kind of interaction to
$N>2$ ions would realize a unitary operation
$U_{MS}^{X}(\theta)=\exp(-i\frac{\theta}{4}{S_x}^2)$ with
$S_x=\sum_{k=1}^N\sigma_x^{k}$, yielding an equal pairwise
coupling between all ions in the string. In addition, using a
single frequency resonant with the qubit transition instead of a
bichromatic field coupling to the transitions' motional sidebands,
the same laser beam could induce spin flips on all ions described
by the unitary $U_X(\theta)=\exp(-i\frac{\theta}{2}{S_x})$. This
opens up the interesting prospect of performing arbitrary unitary
operations by complementing these unitaries by single-qubit phase
shift gates
$U_z^{(k)}(\theta)=\exp(-i\frac{\theta}{2}\sigma_z^{(k)})$ that
could be induced by a tightly focussed off-resonant laser beam
interacting only with the $k$'th ion. In this approach, no
interferometric stability is required between the optical path
lengths of the focussed and the wide beam. Moreover, the use of light shift gates facilitates addressing a single qubit in a string of ion qubits without introducing unwanted state transformations on the neighboring ions since the
phase shift $\theta$ is proportional to the intensity of the laser
field as compared to single-qubit spin flip gates where the
rotation angle $\theta$ is proportional to the field amplitude \cite{OptControlQGates_Footnote0}. In
this way, unitary operations could be realized on a small group of
ions without the need to split and rearrange the ion string
in-between \cite{Rowe2002} and with more modest requirements
regarding the spatial mode profile of the tightly focussed laser
beam.

In this paper, we will use optimal control techniques to find
decompositions of $N$-qubit gates into unitary operations induced
by the set of Hamiltonians
\begin{equation}
\label{eq:setofH}
 {\cal
S}=\{{S_x}^2,S_x,\sigma_z^{(1)},\sigma_z^{(2)},\ldots,
\sigma_z^{(N)}\}\,.
\end{equation}
In contrast to similar applications of optimal control in nuclear
magnetic resonance (NMR) experiments \cite{Khaneja2005} and in
laser-induced femto-chemistry \cite{Tesch:2002,Palao:2003}, we are
interested in the case where only one of the Hamiltonians is
applied at a time. As a consequence, the optimal control algorithm
is required to find decompositions of a given quantum gate by
sequences of laser pulses that interact either with all ions in
the same way or with an individual ion.

The goal of this paper is to find gate decompositions of interest for current state-of-the-art ion trap experiments that are more efficient than decompositions based on gates acting on only one or two qubits at a time.

\section{Basis set of operations}
The mean-field interaction ${S_x}^2$ acting on a string of $N$
ions entangles each ion qubit with each other ion qubit. Thus, on
the one hand, application of this Hamiltonian endows us with the
power to entangle arbitrary pairs of qubits. On the other hand, it
induces entangling interactions that are not always desired. The
situation we are encountering is somewhat similar to the one in
NMR quantum computing where the system Hamiltonian $H_{\rm sys}$
consisting of spin-spin interactions and chemical shifts leads to
a system dynamics that needs to be controlled by radio-frequency
fields interacting with single spins at a time. For this purpose,
techniques have been developed for selectively switching off
certain spin-spin interactions by refocussing techniques
\cite{Vandersypen2004}. In the scenario we envision for the ion
trap system, the role of the radio-frequency fields is taken over
by laser pulses inducing single-qubit gate operations that are
intermittently applied to particular ions.

Refocussing techniques are also applicable to the trapped ion
system. For example, to entangle qubits 1 and 2 in a system of
three qubits, the entangling pulse could be split into two parts
and interleaved with a refocussing pulse to obtain the sequence
$U=U_{MS}^X(\pi/4)U_z^{(3)}(\pi)U_{MS}^X(\pi/4)$. The light shift
pulse on the third qubit flips its phase and effectively reverses
the entangling interactions with qubits 1 and 2. Substituting the
last pulse by its inverse, results in a sequence
$U=U_{MS}^X(-\pi/4)U_z^{(3)}(\pi)U_{MS}^X(\pi/4)$ that entangles
qubits 1 and 2 with qubit 3 without inducing entangling
interactions between 1 and 2.

The set of Hamiltonians (\ref{eq:setofH})
%$\{{S_x}^2,S_x,\sigma_z^{(1)},\ldots,\sigma_z^{(N)}\}$
is sufficient for generating arbitrary unitary operations on a
string of $N$ qubits as can be shown by an explicit construction:
A single-qubit x-rotation acting on qubit $k$ is generated by the
spin echo sequence
% hier die falsche Reihenfolge:
%$U_x^{(k)}(\theta)=U_X(\theta/2)U_z^{(k)}(\pi)U_X(-\theta/2)U_z^{(k)}(-\pi)$,
$U_x^{(k)}(\theta)=U_z^{(k)}(-\pi)\,U_X(-\theta/2)\,U_z^{(k)}(\pi)\,U_X(\theta/2)$.
Arbitrary single-qubit gates can then be performed when combining
this operation with single-qubit phase shift gates
$U_z^{(k)}(\theta)$. Similarly, an operation $U_{MS}^{X-x_k}$ that
entangles all ions with each other except for qubit $k$ is
produced from the $N$-qubit entangling gate $U_{MS}^X(\theta)$ by
the pulse sequence
% wieder falsche Reibenfolge:
%U_{MS}^{X-x_k}=U_{MS}^X(\theta/2)U_z^{(k)}(\pi)U_{MS}^X(\theta/2)U_z^{(k)}(-\pi)$.
$U_{MS}^{X-x_k}=U_z^{(k)}(-\pi)\,U_{MS}^X(\theta/2)\,U_z^{(k)}(\pi)\,U_{MS}^X(\theta/2)$.
Substituting $U_{MS}^X$ in the above sequence by $U_{MS}^{X-x_k}$,
a similar sequence is obtained that entangles all qubits except
for two, and by induction, a two-qubit entangling gate is
constructed between any pair of qubits $m$ and $n$ which together
with arbitrary single qubit gates forms a universal set of gates.

While this construction shows that in principle arbitrary unitary
operations are realizable by pulse sequences generated from the
set $\cal S$, it is of no practical use. For the implementation of
$N$-qubit gate operations, we are interested in finding pulse
sequences that minimize gate errors occurring in ion trap quantum
computing. Therefore, we will be searching for sequences having
either a minimum number of (entangling) pulses or a minimum length
in terms of the sum of pulse angles $\theta_n$ of the individual
pulses.

\section{Optimal control of unitary transformations}

Optimal control techniques have been applied to the problem of
generating specific unitary transformations
\cite{Palao:2003,Khaneja2005,Tesch:2002,Grace:2007} with
applications to systems as different as NMR, neutral atoms in
optical lattices, Josephson junction qubits and trapped ions
\cite{Khaneja2005,Montangero:2007,DeChiara:2008,Timoney:2008}. To
find decompositions of entangling ion trap gates in terms of
pulses generated by Hamiltonians $H_k$ from $\cal S$, we modify a
gradient-ascent algorithm that was developed by Khaneja et al.
\cite{Khaneja2005} in the context of NMR experiments. In their
approach, a unitary transformation $U_{\rm target}$ was searched
for
%
%by constructing a unitary operation $U=\prod_{m=1}^M U_m$ composed
%of the unitaries
%\begin{equation}
%U_m=\exp\left(-i\Delta t\left(H_0+\sum_{k=1}^K u_{km}H_k\right)\right),
%\end{equation}
%
by constructing a unitary operation
\begin{equation}
U=\prod_{m=1}^M U_m=\prod_{m=1}^M e^{\left(-\frac{i}{\hbar}\Delta
t\left(H_{\rm sys}+\sum_{k=1}^K u_{km}H_k\right)\right)},
\label{eq:UKhaneja}
\end{equation}
where $H_k\in {\cal S}$, that maximized the performance function
$\Phi(\{u_{km}\})=|\mbox{Tr}(U^\dagger U_{\rm target})|^2$. The
authors noted that for small time increments $\Delta t$ the
calculation of the gradient \cite{OptControlQGates_Footnote1}
\begin{equation}
\frac{\partial\Phi}{\partial
u_{km}}\approx-2\mbox{Re}(\mbox{Tr}(\frac{i\Delta t}{\hbar}
W_mH_kV_m)\mbox{Tr}(W_mV_m)^\ast), \label{eq:gradient}
\end{equation}
with $W_m=U_{\rm target}^\dagger U_N\cdots U_{m+1}$ and
$V_m=U_m\cdots U_1$, could be efficiently carried out requiring
only about $3M$ matrix multiplications and about $KM$ calculation
of traces. Then, a gradient-based algorithm was devised to
increase the value of the performance function by modifying the
control amplitudes
\begin{equation}
u_{km}\rightarrow u_{km}+\epsilon\frac{\partial\Phi}{\partial
u_{km}}\label{eq:parameterupdate}
\end{equation}
using a suitable step size $\epsilon$. Repeated application of the
gradient calculation followed by updating the control amplitudes
maximized the performance function and resulted in a unitary
transformation realizing the target operation $U_{\rm target}$.

There are a few important differences between coupled spin systems
in NMR and in laser-manipulated strings of trapped ions. In the
NMR context of ref.~\cite{Khaneja2005}, the product of unitaries
%$U=\prod_{m=1}^M U_m$%
in (\ref{eq:UKhaneja}) arises from a discretization of the time
variable that required the approximation of a continuous control
amplitude $u_k(t)$ by a stepwise continuous function with values
$u_{kj}$. Restrictions to the values of $u_k(t)$ are only due to
technical requirements like amplitude or bandwidth limitations of
the radio-frequency equipment used for producing the control
fields. In ion trap experiments, however, limitations exist for
the values that the functions $u_k(t)$ can take on because the
simultaneous application of different control Hamiltonians is
either technically very challenging or physically impossible. In
the former case, the simultaneous application of single qubit
phase shift gates to more than a single ion would require control
of the spatial profile of a laser beam inducing phase shifts by
the ac-Stark effect. More importantly, in the latter case, the
entangling interaction ${S_x}^2$ is produced by an effective
Hamiltonian that precludes the simultaneous application of
single-qubit phase shifts. As a consequence, the control functions
need to satisfy the condition $u_k(t)u_l(t)=0$ for all $k,l$ at
all times $t$. Therefore, the unitary transformation is naturally
decomposed into a product of unitaries and Eq.~(\ref{eq:UKhaneja})
is replaced by
\begin{equation}
U=\prod_{m=1}^M\exp\left(-i\theta_m H_{k_m}\right),
\label{eq:UNebendahl}
\end{equation}
where $\theta_m=\Delta t/\hbar\,u_m$ and $k_m$ labels the
Hamiltonian from set $\cal S$ that is to be used for the {\sl m}th
pulse. As the system is stationary in the absence of laser
interactions, the system Hamiltonian $H_{\rm sys}$ was omitted in
Eq.~(\ref{eq:UNebendahl}).

Using the gradient-ascent method for updating the control
amplitudes $\theta_m$ increases the performance function but
leaves the pulse ordering defined via the indices $k_m$ unchanged.
Thus, for the optimum pulse order to be included in the
configuration space, the number of pulses $M$ needs to be much
larger than the expected minimum number of pulses finally
realizing the target operation $U_{\rm target}$. Therefore, the
search algorithm has to be complemented by a penalty function like $\Phi_p=\sum_{m=1}^M |\theta_m|^\gamma$, $0<\gamma<1$, that tries to eliminate
short pulses that do not contribute much to increasing the
performance function $\Phi$. The functional form of $\Phi_p$
assures that a change in the length $\theta_m$ of the $m$'th pulse
by $d\theta_m$ penalizes already short pulses much more than
longer ones since $d\Phi_p={\rm
sign}{(\theta_m)}\gamma|\theta_m|^{\gamma-1}\, d\theta_m$. In the optimization routine used for finding the pulse decompositions presented in section \ref{sec_examples}, the exponent $\gamma$ ranged from 0.5 to 0.8.
Now, the performance function $\Phi$ is replaced by
$\hat{\Phi}=\Phi-\alpha\Phi_p$ where $\alpha$ is a suitably chosen
weight.

For updating the pulse lengths $\theta_m$, we do not calculate the
gradient of $\hat{\Phi}$ and move in the direction of steepest
ascent but perform consecutive one-dimensional maximizations of
$\theta_m$ instead. This has the advantage that the step size
$\epsilon$ of Eq.~(\ref{eq:parameterupdate}) can be individually
adjusted for the different directions by considering also the
curvatures $\partial^2\hat{\Phi}/\partial \theta_m^2$. For a
negative curvature,the pulse length $\theta_m$ is updated by going
to the maximum of the parabola approximating the $\hat{\Phi}$
whereas for a positive curvature a fixed step size is used for
updating $\theta_m$.

To avoid becoming trapped in a local maximum of the performance
function, we combine the uphill search algorithm with elements of
simulated annealing. Instead of choosing the pulse length
$\theta_m^\ast$ corresponding to the maximum of the parabolic
approximation to $\hat{\Phi}$, the algorithm samples the region
around the maximum by randomly choosing a pulse length
$\theta_m=\theta_m^\ast+\Delta\theta$ where $\Delta\theta$ is
randomly drawn from a normal distribution with probability density
$\propto\exp((\partial^2\hat{\Phi}/\partial
\theta_m^2)(\Delta\theta)^2/T_{\rm eff})$ where the effective
temperature $T_{\rm eff}$ determines the spread of the
distribution around $\theta_m^\ast$. In the course of the
optimization, $T_{\rm eff}$ is lowered to zero. In addition, the
algorithm tries to introduce new pulses into the sequence from
time to time to achieve a variation of the pulse order. Unless
otherwise mentioned the program is started from a random sequence
of pulses of sufficient length.

The computational overhead for performing an update of the pulse
amplitudes is the same for a method following the gradient and for
the $M$ consecutive one-dimensional optimizations of the pulse
lengths. Choosing the one-dimensional optimizations allows us to
calculate the curvatures $\partial^2\hat{\Phi}/\partial
\theta_m^2$ at no additional cost as well as to include the
annealing technique. It would be preferable to use all the
information encoded in the elements $\partial^2\hat{\Phi}/\partial
\theta_m\partial\theta_n$ of the Hesse matrix, however, we judged
the computational cost amounting to ${\cal O}(M^2)$ matrix
multiplications to be prohibitively high.

\section{Examples\label{sec_examples}}

The optimization routine was used to search for decompositions of
unitary transformations of interest in systems of three to five
qubits. In the following subsections, examples will be given for
pulse sequences found by the program. Interestingly, it turns out
that in most cases the sequences consist of pulses having pulse
lengths that are simple fractions of $\pi$ even though the
optimization routine was allowed to vary the pulse lengths
continuously. Moreover, the pulse sequences listed below realize
the desired target operation not only approximately but exactly.

Depending on the initial pulse sequence and the values of the
parameters controlling the optimization process, the optimization
algorithm can converge to different solutions realizing a target
operation $U_{\rm target}$. This demonstrates that we cannot be
sure that the pulse sequences found by the program necessarily
represent the optimum solution. However, as we are interested in
discovering sequences of practical interest to be used in
experiments, this is hardly a drawback.

In the following, we simplify our notation by using the short-hand
notation $U_z^{(k)}(\theta) \leftrightarrow [\theta]_z^k$,
$U_X(\theta) \leftrightarrow [\theta]_X$, $U_{MS}^{X}(\theta)
\leftrightarrow [\theta]_{XX}$ to achieve a convenient and compact
representation of the pulse sequences. Pulses are separated by
hyphens and temporally ordered from left to right. Whenever a
sequence is given, the number of qubits it is operating on is
mentioned in the text.

\subsection{CNOT gates on three qubits}

In a system of three qubits, a CNOT gate between two qubits is
realized by the sequence
\begin{eqnarray*}
&&
[\frac{\pi}{2}]_{X}-[\frac{\pi}{2}]_z^{1}-[\frac{\pi}{4}]_{XX}-
[\frac{\pi}{4}]_{X}-[\pi]_z^{3}-[\frac{\pi}{4}]_{X}-
\\ &&
[\frac{\pi}{4}]_{XX}-[\frac{\pi}{2}]_z^{1}-[\frac{\pi}{2}]_{X}-[\pi]_z^{3},
\end{eqnarray*}
where the target qubit 2 is controlled by qubit 1, corresponding
to the operation $U=({\mathcal
I}+\sigma_z^{(1)}+\sigma_x^{(2)}-\sigma_z^{(1)}\sigma_x^{(2)})/2$.

A unitary transformation consisting of two CNOT operations with
qubit 1 controlling the other two qubits is described by the
unitary operation $U=({\mathcal
I}+\sigma_z^{(1)}+\sigma_x^{(2)}\sigma_x^{(3)}-\sigma_z^{(1)}\sigma_x^{(2)}\sigma_x^{(3)})/2$.
One way of decomposing it into elementary operations is given by
the sequence
\begin{eqnarray*}
&&
[\frac{\pi}{2}]_{X}-[-\frac{\pi}{2}]_z^{1}-[\frac{\pi}{4}]_{XX}-
[-\frac{\pi}{4}]_{X}-[\pi]_z^{1}-[-\frac{\pi}{4}]_{X}-
\\ &&
[-\frac{\pi}{4}]_{XX}-[\frac{\pi}{2}]_z^{1}-[\frac{\pi}{2}]_{X}.
\end{eqnarray*}

As an example of a pulse sequence found by the search algorithm
that is not composed of pulses having pulse angles which are
simple rational fractions of $\pi$, we present another
decomposition of the three-qubit operation $U$ given by \linebreak
% wrong pulse order:
%$[\alpha_1]_z^{1}-[\beta_1]_{YY}-[\alpha_1]_z^{1}-[\beta_2]_{YY}-[\alpha_2]_z^{1}-[\beta_2]_{YY}$.\linebreak
%For $\alpha_1=...$,$\alpha_2=...$, $\beta_1=...$ and $\beta_2=...$,
%
\begin{equation*}
[\beta_2]_{XX}-[\alpha_2]_z^{1}-[\beta_2]_{XX}-[\alpha_1]_z^{1}-[\beta_1]_{XX}-[\alpha_1]_z^{1}.
\end{equation*}
For $\alpha_1\approx 0.7121\pi$, $\alpha_2\approx -0.4241\pi$,
$\beta_1\approx -0.2121\pi$ and $\beta_2\approx 0.3560\pi$, it
also realizes up to an unimportant global phase two CNOT gates on
three qubits with qubit 1 controlling the other two. The program
having provided the pulse angles, we found in a second step that
the angles satisfy the algebraic relations $\alpha_1=2\beta_2$,
$\alpha_2=\pi-4\beta_2$, $\beta_1=\pi/2-2\beta_2$ with
%$\beta_2=\frac{\pi}{2}-\frac{1}{4}\arccos\left(\frac{\sqrt{5}-3}{\sqrt{5}+1}\right)$.
$\beta_2=\frac{3}{8}\pi-\frac{1}{4}\arcsin(\sqrt{5}-2)$.
\subsection{CNOT gates on more than three qubits}
In general, a sequence realizing a two-qubit gate in a system with
$N=3$ qubits will not correctly function in the case $N>3$ as it
risks to entangle the spectator qubits with each other. On the
other hand, a sequence realizing a two-qubit gate for $N=4$ which
does not contain any phase shift gates on the spectator qubits is
also applicable to $N>4$ because of the symmetry of the
interactions between the different spectator qubits. A pulse
sequence realizing a CNOT gate for $N\ge 4$ is given by
\begin{eqnarray}
&&
[\frac{\pi}{2}]_{X}-[-\frac{\pi}{2}]_z^{1}-[-\frac{\pi}{8}]_{XX}-
[\pi]_z^{2}-[\frac{\pi}{8}]_{XX}-
\nonumber\\ &&
[\frac{\pi}{4}]_{X}-[\pi]_z^{1}-[-\frac{\pi}{8}]_{XX}-
[\pi]_z^{2}-[\frac{\pi}{8}]_{XX}- \label{eq:CNOTonNQubits}
\\ &&
[-\frac{\pi}{4}]_{X}-[-\frac{\pi}{2}]_z^{1}-[-\frac{\pi}{2}]_{X}
\nonumber
\end{eqnarray}
where again the qubit 1 controls the target qubit 2.

\subsection{Further multi-qubit operations}
Apart from CNOT gate operations, we also searched for a
decomposition of a quantum Toffoli gate operation. The sequence
\cite{OptControlQGates_Footnote2}
\begin{eqnarray*}
&&
[\frac{\pi}{2}]_{Y}-[\frac{\pi}{4}]_z^{3}-[\frac{\pi}{2}]_{XX}-
[-\frac{\pi}{2}]_{X}-[-\frac{\pi}{2}]_z^{3}-[-\frac{\pi}{4}]_{X}-
\\ &&
[\frac{\pi}{4}]_{XX}-[\frac{\pi}{2}]_z^{3}-[\frac{\pi}{2}]_{XX}-
[\frac{\pi}{2}]_{X}-[-\frac{\pi}{2}]_{Y}
\end{eqnarray*}

is applicable to a system of three qubits and flips the state of
the third qubit depending on the state of qubits 1 and 2, thus
realizing the operation $U=(3{\mathcal
I}+\sigma_x^{(3)}+(\sigma_z^{(1)}+\sigma_z^{(2)}-\sigma_z^{(1)}\sigma_z^{(2)})(\mathcal{
I}-\sigma_x^{(3)}))/4$.

In a system with an even number $N=2M$ of qubits,
%an operation that realizes
a mapping between the Bell basis and the product state basis for
each pair of qubits $(2m-1,2m)$, $m=1,\ldots M$, is of interest as
it could be used for measuring the multipartite concurrence of an
$M$-qubit quantum state available in two copies \cite{Aolita:2006}
and for implementing entanglement purification protocols. For the
case of four qubits, the sequence
\begin{equation*}
[\frac{\pi}{4}]_{XX}-[\pi]_z^{1}-[\pi]_z^{2}-[\frac{\pi}{4}]_{XX}-[\pi]_z^{1}-[\pi]_z^{2}
\end{equation*}
realizes the desired mapping described by
$U_{(1-2,3-4)}=\exp(-i\frac{\pi}{4}(\sigma_x^{(1)}\sigma_x^{(2)}+\sigma_x^{(3)}\sigma_x^{(4)}))$.

The approach can be extended to higher numbers of qubits.
Executing the sequence
\begin{eqnarray*}
&&
[\frac{\pi}{8}]_{XX}-[\pi]_z^{1}-[\pi]_z^{2}-[\pi]_z^{3}-[\pi]_z^{4}-
\\ &&
[\frac{\pi}{8}]_{XX}-[\pi]_z^{1}-[\pi]_z^{2}-[\pi]_z^{5}-[\pi]_z^{6}
\end{eqnarray*}
twice, realizes the unitary transformation
$U_{(1-2,3-4,\ldots)}=\exp(-i\frac{\pi}{4}\sum_{m=1}^M\sigma_x^{(2m-1)}\sigma_x^{(2m)})$
for the case $N=6,8$. Moreover, this approach could also be used
to create linear cluster states and to realize the Hamiltonian of
a 1-D Ising model by concatenating the unitaries
$U_{(1-2,3-4,\ldots)}$ and $U_{(2-3,4-5,\ldots)}$.

\subsection{Quantum error correction}
\begin{figure}[t]
\includegraphics[width=8cm]{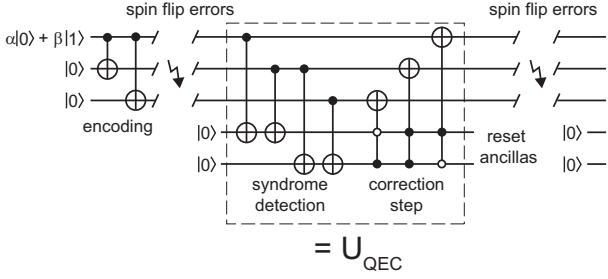}
\caption{\label{fig:errorcorrection} Repetitive error correction
of spin flips using three qubit for encoding the logical qubit and
two ancilla qubits for error syndrome detection and coherent
correction of errors. The optimal control algorithm looked for a
decomposition of the unitary operation enclosed in the dashed box
comprising syndrome detection and error correction.}
%including the syndrome detection, error correction and an
%arbitrary two-qubit unitary gate $V$ on the ancilla qubits. The
%gate $V$ has no influence on the algorithm as the ancillas are
%reinitialized to $|0\rangle$ after its application but it reduces
%the number of constraints the search algorithm has to take into
%account.}
\end{figure}
In experimental quantum information processing, first steps have
been taken to demonstrate quantum error correction
\cite{Cory1998,Chiaverini:2004a} using three-qubit codes. A major
step forward would be the realization of repetitive error
correction where the quantum information remains encoded in a
logical qubit all the time. In the example shown in
Fig.~\ref{fig:errorcorrection}, a qubit state $\alpha|0\rangle +
\beta |1\rangle$ is encoded in a logical qubit consisting of three
qubits as $|\psi_L\rangle=\alpha|000\rangle + \beta |111\rangle$.
The circuit detects single bit flip errors $\sigma_x^{(m)}$,
$m=1,2,3$, by means of two additional ancilla qubits for syndrome
detection and corrects the errors coherently by application of
quantum Toffoli gates in order to restore the state
$|\psi_L\rangle$.

Using gate decompositions similar to Eq.~(\ref{eq:CNOTonNQubits})
for constructing the unitary transformation $U_{\rm QEC}$ that
detects and corrects errors from CNOT and Toffoli gates would
result in a pulse sequence with more than 100 pulses that does not
seem to be realizable with current technology. Therefore, the only
practical approach seems to be to search for a gate decomposition
of the complete operation $U_{\rm QEC}$. However, as the time
needed for finding gate decompositions scales exponentially with
the number of qubits, the task of decomposing the five-qubit
operation $U_{\rm QEC}$ might seem quite challenging at a first
glance. But actually, there is a whole class of operations
$\tilde{U}_{\rm QEC}$ equivalent to $U_{\rm QEC}$ accomplishing
the error correction protocol. To see this, note that we only
require $\tilde{U}_{\rm QEC}$ to perform the mapping
$U^{(m)}|\psi_L\rangle|00\rangle\rightarrow|\psi_L\rangle|\psi_A^{(m)}\rangle$
where
$U^{(m)}\in\{\mathcal{I},\sigma_x^{(1)},\sigma_x^{(2)},\sigma_x^{(3)}\}$
and $|\psi_A^{(m)}\rangle$ are arbitrary orthonormal vectors
describing the ancilla state at the end of the correction step.
Therefore, any valid transformation $\tilde{U}_{\rm QEC}$ can be
put into the form
\begin{eqnarray}
\tilde{U}_{\rm
QEC}P^A_{00}=&\sum_m&\!\!\!\!\left(|000\rangle\langle
000|U_m\otimes|\psi_{A,0}^m\rangle\langle 00|\right.\nonumber\\
&+&\!\!\left.|111\rangle\langle
111|U_m\otimes|\psi_{A,1}^m\rangle\langle 00|\right)
\end{eqnarray}
where the ancilla states needs to satisfy the constraint
$\psi_{A,0}^m=\psi_{A,1}^m$.
%with $\langle\psi_{A,0}|\psi_{A,1}\rangle=1$.
Here, $P^A_{00}={\mathcal I}\otimes|00\rangle\langle 00|$ is a
projector onto the initial ancilla state.
% and $P^L=(|000\rangle\langle 000|+|111\rangle\langle 111|)\otimes
%{\mathcal I}$ projects onto the state space of the logical qubit.
The condition $\langle\psi_{A,0}|\psi_{A,1}\rangle=1$ expressing
the constraint $\psi_{A,0}^m=\psi_{A,1}^m$ assures that no
information about the state of the logical qubit can be obtained
from detecting the ancilla state. To search for a gate decomposition, the performance function can
now be modified by replacing the trace $\mbox{Tr}(U^\dagger U_{\rm
QEC})$ by
$\Phi=\mbox{Re}(\sum_m\langle\psi_{A,0}^{(m)}|\psi_{A,1}^{(m)}\rangle)$.

Using this approach, the search program found the following pulse
decomposition $\tilde{U}_{\rm QEC}$ for realizing an operation equivalent to $U_{\rm QEC}$:
\begin{widetext}
\begin{eqnarray*}
&&
[-\frac{\pi}{2}]_{X}-[\frac{\pi}{2}]_z^{5}-[\frac{\pi}{2}]_z^{4}-
[\pi]_z^{3}-[-\frac{\pi}{8}]_{YY}-[\pi]_z^{2}-[-\frac{\pi}{8}]_{YY}-
[\pi]_z^{5}-[\pi]_z^{3}-[\frac{\pi}{8}]_{YY}-[\pi]_z^{2}-[-\frac{3\pi}{8}]_{YY}-[\frac{\pi}{2}]_{X}-[\frac{\pi}{2}]_z^{3}-
\\
&&
[\frac{\pi}{2}]_z^{1}-
[-\frac{\pi}{8}]_{XX}-[\pi]_z^{5}-[\pi]_z^{4}-[\frac{\pi}{8}]_{XX}-[\pi]_z^{1}-
[\frac{\pi}{2}]_z^{4}-[\pi]_z^{5}-[\frac{\pi}{8}]_{XX}-[\pi]_z^{2}-[-\frac{3\pi}{8}]_{X}-[\pi]_z^{4}-[-\frac{\pi}{8}]_{XX}-[\pi]_z^{5}-
\\
&&
[\frac{\pi}{8}]_{XX}-[\frac{\pi}{8}]_{X}-[\pi]_z^{2}-[\frac{\pi}{8}]_{XX}-
[-\frac{\pi}{2}]_z^{4}-[-\frac{\pi}{4}]_{XX}.
\end{eqnarray*}
\end{widetext}
Here, the uphill search was extremely slow when starting the
search algorithm from a random sequence of pulse. However,
starting the algorithm from a sequence constructed from gate
decompositions of the CNOT and Toffoli gates and initially driving
it away from this undesired solution by increasing the effective
temperature $T_{\rm eff}$ proved to be an effective strategy for
finding an improved solution.

\section{Summary and outlook}
In conclusion, we have developed an algorithm based on optimal
control techniques for finding decompositions of unitary
transformations into finite sequences of pulses that correspond to
the application of Hamiltonians drawn from a given set. For a set
of Hamiltonians that are of interest for ion trap quantum
computing, the algorithm provides gate decompositions that would
be difficult to find otherwise.

Looking for gate decompositions involving parallel two-qubit interactions
on a small number of ions provided in many cases (as for example for the quantum Toffoli gate) a pulse sequence much shorter than what would have been possible by using sequential two-qubit interactions and single qubit gates. In a few other cases, however, this strategy did not pay off (as we noted when looking for a decomposition of the quantum Fredkin gate).

While our investigation was limited to a particular basis set of
Hamiltonians $\cal S$, the program could be adapted to search for
gate decompositions using other sets that might be more relevant
for realizations of quantum computing in other physical systems
where, for example, the natural entangling interaction is given by
a $\sqrt{i\rm SWAP}$ gate \cite{Steffen2006a} or an exchange
interaction \cite{DiVincenzo2000}. For example, in the context of
ion trap quantum computing, another interesting question is
whether the algorithm could be modified to make it applicable to
the case of quantum gates realized by a tightly focussed laser
interacting with a single ion at a time. In this case, the Hilbert
space would be comprised not only of the qubit states but also of
the harmonic oscillator the qubits are coupling to, a
configuration not only found in ion trap quantum computing but
also in cavity-QED setups where an electromagnetic field mode is
interacting with cold atoms or superconducting flux qubits.
\newline

\begin{acknowledgments}
We gratefully acknowledge the support of the European network
SCALA, the Institut f{\"u}r Quanteninformation GmbH and IARPA.
C.~F.~R. would like to thank T.~Schulte-Herbr{\"u}ggen and
S.~Glaser for useful discussions and C.~Kruszynska for help with
symbolic calculations.

\end{acknowledgments}

\bibliographystyle{apsrev}
%------------------------------------------

%\bibliography{biblio-most-recent}

\begin{thebibliography}{27}
\bibitem[{Opt({\natexlab{a}})}]{OptControlQGates_Footnote0}
\bibinfo{note}{Assuming that a single-qubit gate is performed with a laser beam
  with intensity $I$ and a Gaussian spatial beam profile with waist $w$, it is
  advantageous to work with an interaction whose strengths falls off as $\theta
  (r)\propto\exp(-2r^2/w^2)$ instead of $\theta
  (r)\propto\sqrt{I}\propto\exp(-r^2/w^2)$.}
\bibitem[{Opt({\natexlab{b}})}]{OptControlQGates_Footnote1}
\bibinfo{note}{If the sum in eq.~(\ref{eq:UKhaneja}) consists of only a single
  term, then eq.~(\ref{eq:gradient}) becomes exact.}
\bibitem[{Opt({\natexlab{c}})}]{OptControlQGates_Footnote2}
\bibinfo{note}{As the Hamiltonians in set $\cal S$ are sufficient for
  generating arbitrary unitary operations, we did not include the Hamiltonians
  $S_Y$, $S_Y^2$. However, as these Hamiltonians are experimentally as easily
  generated as $S_X$, $S_X^2$ by shifting the optical phase of the control
  laser field by $\pi/2$ with the help of an acousto-optical modulator, we use
  them in some of the pulse sequences for constructing the desired target
  operation.}
\end{thebibliography}

%------------------------------------------

%------------------------------------------

\end{document}